\documentclass[twocolumn,preprintnumbers,superscriptaddress,aps,prd,floatfix,nofootinbib]{revtex4-2}

\usepackage{enumerate}
\usepackage{amsmath}
\usepackage{graphicx}
\usepackage{placeins}
\usepackage{xspace,slashed}
\usepackage{hyperref}
\usepackage{braket}
\hypersetup{colorlinks=true, citecolor=blue, urlcolor=blue, linkcolor=blue}
\usepackage[normalem]{ulem}
\usepackage{booktabs}
\usepackage{multirow}
\usepackage{orcidlink}

\usepackage[capitalize]{cleveref}
\usepackage{siunitx}

\newcommand{\program}[1]{\texttt{#1}\xspace}

\newcommand{\eos}{\program{EOS}}
\newcommand{\smelli}{\program{smelli}}
\newcommand{\version}[1]{\texttt{v#1}\xspace}

\newcommand{\hc}{\mathrm{h.c.}}

\usepackage{suffix} % Allows us to define a starred command
\newcommand{\V}[1]{\ensuremath{V_{#1}^{}}} 
\WithSuffix\newcommand\V*[1]{\ensuremath{V_{#1}^*}}
\newcommand{\yd}[1]{\ensuremath{y^d_{#1}}\xspace}
\WithSuffix\newcommand\yd*[1]{\ensuremath{y^{d*}_{#1}}\xspace}
\newcommand{\yu}[1]{\ensuremath{y^u_{#1}}\xspace}
\WithSuffix\newcommand\yu*[1]{\ensuremath{y^{u*}_{#1}}\xspace}

\newcommand{\deltapc}{\ensuremath{\delta_\textrm{pc}}\xspace}
\newcommand{\DeltaBSM}{\ensuremath{\Delta^\textrm{BSM}}\xspace}

\begin{document}

\preprint{IPPP/26/36}
\preprint{SI-HEP-2026-11}
\preprint{P3H-26-032}

%%%%%%%%%%%%%%%%%%%%
\title{New insights into the $b\rightarrow c \bar{u}q$ puzzle through Top-Bottom synergies}
%%%%%%%%%%%%%%%%%%%%
\begin{abstract}
Anomalies in the non-leptonic $\bar{B}^0\rightarrow D^{(*)+}K^{(*)-}$ and $\bar{B}^0_s\rightarrow D^{(*)+}_s\pi^-$ decays may be an indication of physics beyond the Standard Model, but the large deviations require strongly coupled new physics that should be visible at colliders.
We explore three new directions that could lead to viable new physics models, performing a detailed collider study to examine the possible weakening of previously known constraints on additional $SU(2)_L$ doublets.
Our results show that, despite the difficulty of probing $t\bar{t}$ final states, increasing the branching ratio to this decay mode does not significantly weaken the bounds on weak doublet scalars, as additionally existing charged Higgs searches are equally strong.
Beyond this, we analyse a potentially large breakdown of QCD factorisation by including large-power corrections to $B$ decays, and the effect of diluting collider searches with multi-scalar extensions.
We find that these typical model-building routes for constructing a viable scenario remain constrained by collider measurements, indicating that these non-leptonic anomalies remain among the most puzzling discrepancies from the SM.
\end{abstract}

\author{Jack Y. Araz\orcidlink{0000-0001-8721-8042}} \email{j.araz@ucl.ac.uk}
\affiliation{Department of Physics and Astronomy, University College London, London, WC1E 6B, UK\\[0.10cm]}
\affiliation{Department of Engineering, City St.\ George’s, University of London, London, EC1V 0HB, UK\\[0.10cm]}
%%%%
\author{Christoph Englert\orcidlink{0000-0003-2201-0667}} \email{christoph.englert@manchester.ac.uk}
\affiliation{Department of Physics \& Astronomy, University of Manchester, Oxford Road, Manchester M13 9PL, UK\\[0.1cm]}
%%%%
\author{Matthew Kirk\orcidlink{0000-0003-0845-7227}} \email{matthew.j.kirk@durham.ac.uk}
\affiliation{Institute for Particle Physics Phenomenology, Durham University, Durham, DH1 3LE, United Kingdom\\[0.10cm]}
%%%%
\author{Gilberto Tetlalmatzi-Xolocotzi\orcidlink{0000-0001-8083-0335}} \email{gtx@physik.uni-siegen.de}
\affiliation{Physik Department, Universit\"{a}t Siegen, Walter-Flex-Str. 3, 57068 Siegen, Germany\\[0.10cm]
}
%\affiliation{Universit\`e Paris-Saclay, CNRS/IN2P3, IJCLab, 91405 Orsay, France\\[0.1cm]}
%%%%
%%%%%%%%%%%%%%%%%%%%%%%%%%%%%%%%%%%%%%%%%%%%%%%%%%%%%%%%%%%%%%%%%

%\flushbottom
\allowdisplaybreaks
%%%%%%%%%%%%%%%%%%%%%%%
%%%%%%%%%%%%%%%%%%%%%%%
\maketitle
%%%%%%%%%%%%%%%%%%%%%%%%%%%%%%%%%%
%%%%%%%%%%%%%%%%%%%%%%%%%%%%%%%%%%
\section{Introduction}
\label{sec:intro}
%%%%%%%%%%%%%%%%%%%%%%%%%%%%%%%%%%
Non-leptonic $B$ meson decays are of great relevance for multiple reasons; for instance, they allow us to probe $CP$ violation and test the CKM mechanism of quark flavour mixing. Currently, there is ample experimental evidence achieving precision below $3\%$ in some cases. Thus, it is crucial to have robust theoretical tools that enable us to leverage the available experimental results. One of the main issues with these computations is the presence of hadronic contributions with large uncertainties, which severely affect most of these processes.  A widely accepted formalism for calculating these decays is QCD factorisation (QCDF) \cite{Beneke:1999br}. In general, the presence of weak annihilation and penguin topologies limits the precision achievable in the corresponding theoretical calculations.

However, the study of the processes $\bar{B}^0\rightarrow D^{(*)+}K^{(*)-}$ and $\bar{B}^0_s\rightarrow D^{(*)+}_s\pi^-$ presents a clean alternative, since the absence of an equal flavour quark-antiquark pair in the final state mesons leads to the absence of annihilation topologies. This is an enormous advantage, since they are the source of infrared divergences that currently cannot be addressed from first principles and thus, in general, occupy a large fraction of the error budget. Moreover, within the SM, these decays are described by pure tree-level topologies arising from the quark-level transitions $b \rightarrow c \bar{u} d$ and $b \rightarrow c \bar{u} s$ (which we will refer to generically as $b \rightarrow c \bar{u} q$).

A recent update on these decay channels found sizeable discrepancies between theoretical predictions and experimental determinations of these decays \cite{Bordone:2020gao}, with tensions that can exceed five standard deviations, giving rise to the so-called ``$b\rightarrow c \bar{u}q$ puzzle''.
Two possible explanations have been suggested in the literature: the first is an underestimation of the hadronic uncertainties \cite{Endo:2021ifc, Meiser:2024zea}; the second is that these large deviations between theory and experiment are indirect signs of new physics (NP) \cite{Iguro:2020ndk, Cai:2021mlt, Bordone:2021cca} \footnote{The possibility of having NP in other non-leptonic $B$ meson decays has been discussed systematically using effective field theory methods in \cite{Bobeth:2014rra, Bobeth:2014rda,Brod:2014bfa,Lenz:2019lvd,Alguero:2020xca, Biswas:2023pyw,Biswas:2024bhn,Biswas:2025drz, Fleischer:2021cct, Fleischer:2021cwb}.}. In the absence of additional insights, the possible explanation remains ambiguous.

Regarding an underestimation of theoretical uncertainties, we first reiterate that within QCDF, the corrections to the processes $\bar{B}^0\rightarrow D^{(*)+}K^{(*)-}$ and $\bar{B}^0_s\rightarrow D^{(*)+}_s\pi^-$ beyond leading order in $\Lambda_{\rm QCD}/m_b$ are expected to be small. However, since we cannot address these corrections from first principles, it cannot be ruled out that these effects are larger than these first estimations. A reasonable question is then, how large can these extra purely QCD effects really be?

One way to assess the feasibility of the NP explanation is to analyse the implications of the beyond the Standard Model (BSM) effects in the $b\rightarrow c \bar{u}q$ transitions in other sectors of high-energy physics. At the end of the day, the NP mediators entering in  $b\rightarrow c \bar{u}q$ should also affect other measurements, such as collider processes. This, however, requires a mechanism to mediate the interplay between low-energy interactions and high-energy collider decays. Such an approach has been followed, for instance, by Refs.\ \cite{Bordone:2021cca} and \cite{Atkinson:2024hqp}. In particular, Ref.\ \cite{Bordone:2021cca} found that dijet searches challenge the NP hypothesis at low energies.\footnote{In that work, a narrow window of viability for an additional Higgs doublet with a mass very close to that of the top quark was identified -- this idea will not be pursued further in this work.}

One of the main issues of these approaches is that by assuming the minimal couplings that directly link $B$-meson physics and the collider phenomenology, other decay paths that could alter the collider picture are neglected. In particular, $t \bar{t}$ final states are much less sensitive to BSM effects (see e.g.~\cite{ATLAS:2024vxm,CMS:2025dzq}), so models where this is the dominant mode could see significantly different collider bounds.

In this work, we consider three scenarios that would affect our understanding of potential NP in the selected non-leptonic $B$ decays:
\begin{enumerate}
    \item NP is hiding in discovery modes that are difficult to constrain, namely through top-philic components that are systematics-limited due to large, accidental signal-background interference~\cite{Gaemers:1984sj}. 
    \item QCD is not as well under control as we believe and creates tension in the $B$ sector through unexpectedly large power corrections.
    \item The NP sector contains multiple new particles, which dilutes collider bounds (e.g.\ nHDM for \hbox{$n > 2$}).
\end{enumerate}
We set out to discover how these three scenarios, and their interplay, affect the viability of a specific NP hypothesis, namely a two-Higgs-doublet model (2HDM).

%%%%%%%%%%%%%%%%%%%%%%%%%%%%%%%%%%
\section{Overview of New Physics in Non-leptonic B meson decays}
%%%%%%%%%%%%%%%%%%%%%%%%%%%%%%%%%%
As discussed, the processes $\bar{B}^0\rightarrow D^{(*)+}K^-$ and $\bar{B}^0_s\rightarrow D^{(*)+}_s\pi^-$ are particularly clean, and these form the basis of our low-energy analysis, based on our previous work in \cite{Atkinson:2024hqp}. In QCDF, the relevant decay amplitudes can be written as\footnote{The decay channels $\bar{B}^0\rightarrow D^{(*)+}K^-$ and $\bar{B}^0_s\rightarrow D^{(*)+}_s\pi^-$ belong to the so-called Class-I processes \cite{Neubert:1997uc}, see e.g.\ \cite{Fleischer:2010ca} for an early phenomenological study.} \cite{Beneke:2000ry}
\begin{multline}
\label{eq:amplitude_with_power_corrections}
\mathcal{A}\left( \bar{B}_{(s)}^{0} \to D_{(s)}^{(*)+} L^{-} \right) \\= A_{D_{(s)}^{(*)+} L^{-}} \left[  a_1 \left( D_{(s)}^{(*)+} L^{-} \right) + \deltapc \right] \,,
\end{multline}
where the light meson is $K^-$ for an initial state $\bar{B}$ and $\pi^-$ for an initial state $\bar{B}_s$.
Besides the relevant CKM elements, the coefficient $A_{D_{(s)}^{(*)+} L^-}$ includes the form factor $F_j^{B \to D^{(*)}}$ for the transition $B \to D^{(*)}$ and the decay constant $f_L$ of the corresponding light meson. The sub-amplitude $a_1\left( D_{(s)}^{(*)+} L^- \right)$ is calculated perturbatively, and currently is known up to NNLO in the \( \alpha_s \) expansion~\cite{Huber:2016xod} for the SM, while for all possible BSM effective four-quark operators the relevant hard-scattering kernels have been determined up to NLO within QCDF in Refs.\ \cite{Cai:2021mlt,Meiser:2024zea}. The parameter \deltapc accounts for any possible non-factorisable extra effects suppressed by powers in $\Lambda_{\rm QCD}/m_b$, which for the clean decays we are considering should be small under QCDF assumptions, with current estimates at the level of $1\%$~\cite{Bordone:2020gao}.\footnote{Alternative estimations within the context of QCD sum rules are also available, reporting large uncertainties \cite{Piscopo:2023opf}.} In \cref{sec:power-corrections}, we discuss in more detail the possibility that the power corrections have so far been significantly underestimated.

To improve our theoretical predictions, we work with the ratios
\begin{equation}
\begin{split}
	R_{(s)L}^{(\ast)} &\equiv \frac{\Gamma(\bar{B}_{(s)}^0\to D_{(s)}^{(\ast)+}L^-)}{\text{d}\Gamma(\bar{B}_{(s)}^0\to D_{(s)}^{(\ast)+}\ell^-\bar{\nu}_{\ell})/\text{d}q^2\mid_{q^2=m_L^2}}\\\ 
	&=6\pi^2\,|V_{uq}|^2\,f_L^2\,|a_1(D_{(s)}^{(\ast)+}L^-)|^2\, X_{(s)L}^{(\ast)}\,.
\label{eq:ratios}	
\end{split}
\end{equation}
where $X_L$ factors are ratios of form factors (the general definition can be found in~\cite{Neubert:1997uc} for pseudoscalar and vector mesons $D^{(*)}$).
This construction ensures that the $V_{cb}$ dependence vanishes (which is helpful in light of ongoing tensions between inclusive and exclusive determinations; see the PDG review~\cite{ParticleDataGroup:2024cfk,PDG:2024VcbVub} for a summary) and reduces the form-factor dependence. Numerically, we evaluate these form-factor ratios using the software \eos~\cite{EOSAuthors:2021xpv}, which enables us to incorporate state-of-the-art results while accounting for correlations. The leading uncertainties in this ratio now stem from the decay constant $f_L$ and the renormalisation scale uncertainty, which affects $a_{1}(D^{(*)+}_{(s)}L^{-})$, both of which are at the per cent level. Our theoretical predictions can be seen in \cref{tab:ObsR}.

%%%%%%%%%%%%%%%%%%%%%%%%%%%%%%%%%%
\begin{table}
\renewcommand{\arraystretch}{1.7}  % Increase line spacing
\begin{tabular}{c c c c c}
\toprule
Transition & Observable & Experiment & SM & Pull \\
\midrule
\multirow{2}{*}{$b \to c\bar{u}d$} & $R_{s\pi}$     & $0.71\pm 0.06 $            & $1.06^{+0.04}_{-0.03}$    & $\approx \qty{5}{\sigma}$
\\
                      & $R^{*}_{s\pi}$ & $0.52^{+0.18}_{-0.16}$     & $1.05^{+0.04}_{-0.03}$    & $\approx \qty{3.1}{\sigma}$
\\
\addlinespace
\multirow{3}{*}{$b \to c\bar{u}s$} & $R_K$          &  $0.058^{+0.004}_{-0.004}$ & $0.082^{+0.002}_{-0.001}$ & $\approx \qty{5.6}{\sigma}$
\\ 
                      & $R_{K^*}$      & $0.136\pm 0.023 $          & $0.14^{+0.01}_{-0.01}$    & $\approx \qty{0.16}{\sigma}$
\\
                      & $R^*_{K}$      & $0.064\pm 0.003$           & $0.076^{+0.002}_{-0.001}$ & $\approx \qty{3.6}{\sigma}$
\\
\bottomrule
\end{tabular}
\caption{Observables used in our low-energy $B$-physics analysis, and the discrepancy between SM and experiment.\label{tab:ObsR}}
\end{table}
%%%%%%%%%%%%%%%%%%%%%%%%%%%%%%%%%%

%%%%%%%%%%%%%%%%%%%%%%%%%%%%%%%%%%
\section{New Physics models for non-leptonic decays}
%%%%%%%%%%%%%%%%%%%%%%%%%%%%%%%%%%
Previous work \cite{Bordone:2021cca} has studied single field extensions of the SM, using the minimal couplings needed to produce the desired effect in the low-energy non-leptonic $B$ decay observables. Amongst others, the authors of \cite{Bordone:2021cca} studied a 2HDM in the alignment limit where the second doublet couples only to the left-hand quark doublet and right-handed down quarks (benchmark 1 in ``Colourless scalar doublet model'', hereafter BGM), with the BSM Lagrangian:
\begin{equation}
\mathcal{L}_\text{BGM} \supset -\yd{ij} \overline{Q_{L,i}} d_{R,j} \Phi + \hc
\end{equation}
with $Q_{L,i} = (V^*_{ji} u_{L,j} \; d_{L,i})^T$ and $\Phi = (-i H^+, (H^0 + i A^0)/\sqrt{2})^T$.
They take $y^d_1 = y^d_2 \neq 0$, $y^d_3 \neq 0$.\footnote{The coupling $y^d$ in BGM is related to our coupling by $y^d|_\textrm{BGM} = -(y^d)^*|_\textrm{this work}$ }
Their study concluded that dijet limits ruled out all the parameter space that could explain the non-leptonic anomaly.

For doublet masses above the $t\bar t$ threshold, a strong collider constraint can arise from the neutral component of the doublet decaying into $t\bar t$. In fact, a set of exotic states in scalar extensions will be top-philic, as they parameterise an orthogonal direction relative to the eaten Goldstone bosons, thereby giving mass to the SM's massive weak gauge bosons. It is known that these searches suffer from significant interference from QCD backgrounds~\cite{Gaemers:1984sj}. 
Thus, if we extend the BGM coupling model by
\begin{equation}
\label{eq:bsm_lagrangian}
\mathcal{L}_\text{BSM} = -\yd{ij} \overline{Q_{L,i}} d_{R,j} \Phi - \yu{ij} \overline{Q_{L,i}} u_{R,j} \tilde{\Phi} + \hc \,,
\end{equation}
allowing couplings to RH up-type quarks, current constraints could be significantly weakened as they are dominated by systematic uncertainties~\cite{ATLAS:2024vxm,CMS:2025dzq} associated with the $t\bar t$ resonance distortion. This will enable us to a priori decouple the electrically neutral scalar phenomenology from its charged counterparts. We can then clarify whether there exists a perturbatively reasonable parameter range in which neutral scalar decay to top final states could be phenomenologically dominant but obscured, hiding the resonance due to interference with other exotic searches that are not yet sensitive enough to claim discovery.

Following BGM, to avoid tree level FCNC in $B$ and $K$ mesons we choose $\yd{}$ to be diagonal (i.e.\ $\yd{} = \textrm{diag} (\yd{1}, \yd{2}, \yd{3})$).
After electroweak symmetry breaking, the charged and neutral components get the following interactions:
\begin{multline}
\mathcal{L}_\textrm{charged} \supset +i H^+ \left( (V.\yd{})_{ij} \bar{u}_{L,i} d_{R,j} + \yu*{ij} \bar{u}_{R,j} d_{L,i} \right) \\+ \hc\,,
\end{multline}
and
\begin{multline}
\mathcal{L}_\textrm{neutral} \supset -\frac{H^0 - i A^0}{\sqrt{2}} \left( \yd{ij} \bar{d}_{L,i} d_{R,j} + (V.\yu{})_{ij} \bar{u}_{L,i} u_{R,j} \right) \\+ \hc
\end{multline}
Later, we will allow a non-zero value for \yu{33}, which opens up the $H^0 \to t\bar{t}$ decay mode, in order to assess the effects on the collider bounds (while having no effect on the low energy non-leptonic decays).

%%%%%%%%%%%%%%%%%%%%%%%%%%%%%%%%%%
\subsection*{Explaining the non-leptonic anomalies}
%%%%%%%%%%%%%%%%%%%%%%%%%%%%%%%%%%
Taking the NP Lagrangian above, a minimal model for the non-leptonic $B$ anomalies only requires $\yd{3} \neq 0$ and $\yd{1} \textrm{ or } \yd{2} \neq 0$.
We follow BGM in setting $\yd{2} = \yd{1}$, since this is favoured by the data (when assuming no power corrections).
Integrating out the doublet at a scale $\mu$, and then taking into account the leading log QCD running, gives low-energy Wilson coefficients:
\begin{equation}
\label{eq:eft_wcs}
\begin{aligned}
C_2^{SRL, \textrm{bcud}} (\mu_\textrm{low}) &= C_2^{SRL, \textrm{bcus}} (\mu_\textrm{low}) 
\\
&= \frac{-\sqrt{2}}{4 G_F} \frac{\yd{3} \yd*{1}}{M_{H^+}^2} \left( 1 + 4 \frac{\alpha_s}{\pi} \ln \frac{\mu}{\mu_\textrm{low}} \right) \,,
\end{aligned}
\end{equation}
which appear in the following low-energy effective Lagrangian
\begin{equation}
\mathcal{L} \supset \frac{4 G_F}{\sqrt{2}} \V{cb} \sum_{q=d,s} \V*{uq} C_2^{SRL, \textrm{bcuq}} (\bar{c}_\alpha P_R b_\alpha)(\bar{q}_\beta P_L u_\beta) \,,
\end{equation}
where $\alpha, \beta$ are colour indices. This matches the Lagrangian defined in \cite{Cai:2021mlt}, except that we explicitly separate the two flavour structures. The QCD running enhances the size of the BSM effect: running down to $\mu_\textrm{low} = \qty{5}{\GeV}$ results in a numerical RG factor of approximately \numlist{1.7;1.8;1.85} for $\mu =$ \qtylist{0.5;1;1.5}{\TeV} respectively.

As in \cite{Atkinson:2024hqp}, we construct a $\chi^2$ function from the non-leptonic data in \cref{tab:ObsR}, and minimise this to find the BSM interaction that can explain the observed discrepancy. Since our NP model enters the amplitude directly at tree level, we can simply express the best fit point as:
\begin{equation}
\label{eq:non-leptonic_best_fit}
\DeltaBSM \equiv \frac{\yd{3} \yd{1}}{M_{H^+}^2} \times \left( 1 + 4 \frac{\alpha_s}{\pi} \ln \frac{M_{H^+}}{M_Z} \right) \approx \frac{-3.55}{\mathrm{TeV}^2} \,.
\end{equation}
A BSM effect of this size reduces the $\chi^2$ by 61, corresponding to a pull of \qty{7.8}{\sigma} relative to the SM. This is our baseline scenario, and in what follows, we study whether such a model (or minimal extensions thereof) can be made compatible with collider data.

%%%%%%%%%%%%%%%%%%%%%%%%%%%%%%%%%%
\section{Signatures at LHC}
\label{sec:collider}
%%%%%%%%%%%%%%%%%%%%%%%%%%%%%%%%%%
%
%%%%%%%%%%%%%%%%%%%%%%%%%%%%%%%%%%
\begin{figure*}[!t]
    \centering
    \includegraphics[width=\linewidth]{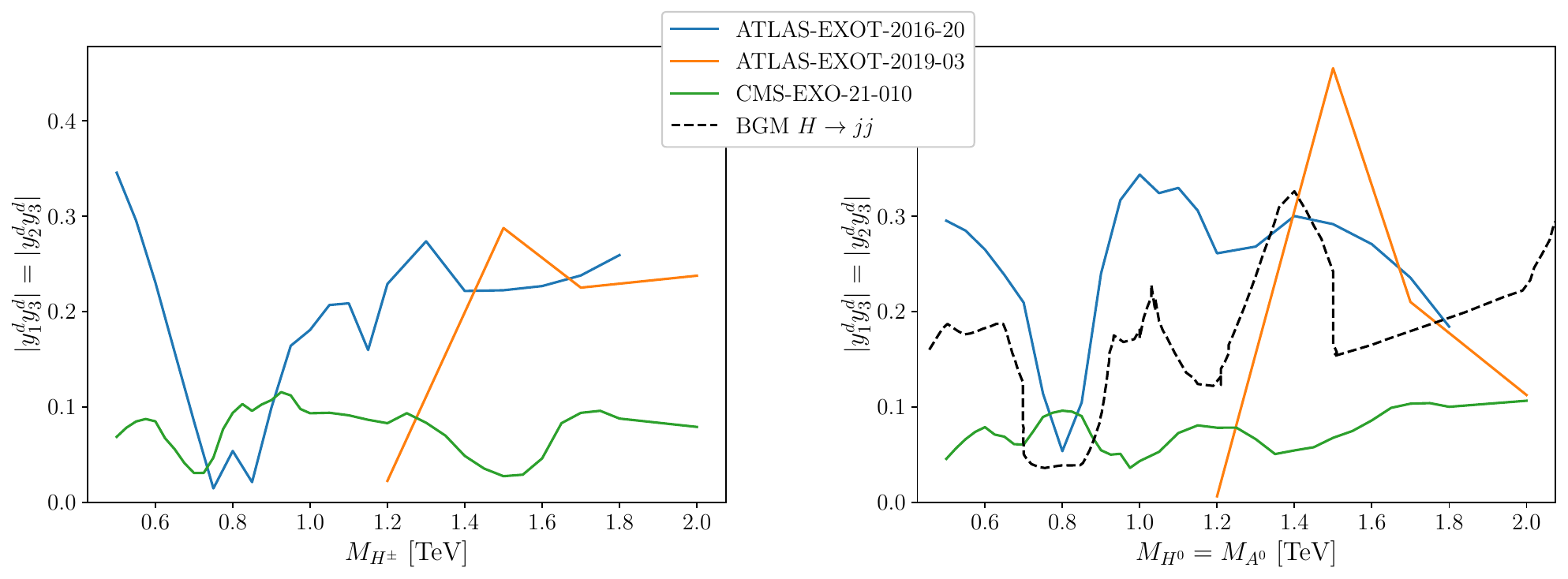}
    \caption{Bounds on the couplings of the charged (left) and CP-even neutral (right) components of the extra doublet from our recasting of ATLAS and CMS results, along with limits found in BGM \cite{Bordone:2021cca}.}
    \label{fig:collider_bounds_individual}
\end{figure*}
%%%%%%%%%%%%%%%%%%%%%%%%%%%%%%%%%%
%
To isolate the effects of the Yukawa couplings and the physical charged or scalar Higgs mass from those of the remaining 2HDM spectrum, all additional Higgs components are decoupled by taking their masses to sufficiently large values. We then perform a scan over the mediator mass and the Yukawa couplings $\yd{1}$, $\yd{3}$, and $\yu{33}$. Results are presented as a function of the product $|\yd{1} \yd{3}|$, with $\yu{33}$ fixed to a benchmark value drawn from the set $\{0, 0.5, 1\}$. The coupling $\yd{1}$ is scanned over the range $[0.01, 0.8]$ and $\yd{3}$ over $[0.01, 1.5]$. The exclusion boundary is then defined as the maximum value of $|\yd{1} \yd{3}|$ consistent with the observed cross-section upper limits at each mediator mass value.

Throughout this study, collider bounds are implemented via comparison against a set of analyses published by the ATLAS and CMS collaborations. Specifically, we use the observed cross-section upper limits provided by ATLAS-EXOT-2016-20~\cite{ATLAS:2018qto}, CMS-EXO-16-056~\cite{CMS:2018mgb}, CMS-EXO-21-010~\cite{CMS:2020zti}, and ATLAS-EXOT-2019-03~\cite{ATLAS:2019fgd}, all of which target dijet final states and span a wide range of dijet invariant masses, providing broad coverage of the signal parameter space considered~here.

Signal events were generated, and partonic cross-sections computed using \textsc{MadGraph5\_aMC@NLO} (version~3.5.11)~\cite{Alwall:2014hca}. Parton distribution function (PDF) effects were included via \textsc{LHAPDF6}~\cite{Buckley:2014ana} using the \textsc{NNPDF 4.1} set~\cite{NNPDF:2021njg}. For the primary, naive cross-section comparison, a standard set of particle-level requirements was applied: all jets were required to carry transverse momentum $p_T > 100$ GeV, with the leading jet satisfying $p_T > 200$ GeV; the dijet invariant mass was required to exceed $m_{jj} > 450$ GeV; all jets were restricted to the pseudorapidity range $|\eta| < 2.8$; and the angular separation between the two leading jets was required to satisfy $0.4 < \Delta R < 1.2$. A parameter point is then excluded wherever the predicted signal cross-section, evaluated after these acceptance requirements, exceeds the observed upper limit provided by the most constraining analysis at that point.

To assess the reliability of this procedure, we additionally performed a full recast of ATLAS-EXOT-2019-03 within the \textsc{MadAnalysis 5} (version~2.0.9) framework~\cite{ATLAS:2019fgd, Conte:2018vmg, Dumont:2014tja, Conte:2014zja, Conte:2012fm}, using the simplified fast-simulation (SFS) interface~\cite{Araz:2020lnp}. In this case, partonic events were passed to \textsc{Pythia 8}~\cite{Bierlich:2022pfr} for parton showering and hadronisation, without the particle-level preselection described above. Exclusion limits were then computed from the resulting signal yields using the most sensitive signal region at each parameter point, with statistical inference performed via the \textsc{Spey} package (version~0.2.6)~\cite{Araz:2023bwx, araz_2025_17853562}.

The recast yields slightly looser bounds compared to the naive cross-section comparison, particularly at low mediator masses, where acceptance and efficiency corrections have the greatest impact. Since full recasts are not available for all four analyses, and given that the naive comparison provides consistently more conservative exclusions, we adopt the naive cross-section limits as our primary collider constraint throughout. This choice does not qualitatively affect our conclusions, as the parameter regions of interest are independently constrained by additional complementary probes.

It is important to note that we observed that CMS-EXO-16-056 is unable to exclude large portions of the scalar Higgs mass spectrum, due to the limited parameter ranges we chose for the Yukawa couplings. However, since CMS-EXO-21-010 is typically the strongest bound (see \cref{fig:collider_bounds_individual}; additional details concerning the systematics of our reparameterisation can be found in~\cref{sec:app}), we did not extend our coupling limits to explore the exact bounds originating from CMS-EXO-16-056. 

On the right plot in \cref{fig:collider_bounds_individual}, we additionally compare our results to those found in \cite{Bordone:2021cca}, which reinterpreted three of the same searches as this work (ATLAS-EXOT-2016-20, ATLAS-EXOT-2019-03, CMS-EXO-16-056) alongside CMS-EXO-19-012. Since our bounds primarily come from CMS-EXO-21-010, which postdates Ref.~\cite{Bordone:2021cca}, we would not expect them to match those in \cite{Bordone:2021cca}; indeed, the new results from CMS-EXO-21-010 are even stronger. We use the best result at each mass value (also assuming no splitting between charged and neutral components, as this is strongly constrained by electroweak precision tests) to construct a ``best'' collider bound going forward.

%%%%%%%%%%%%%%%%%%%%%%%%%%%%%%%%%%
\begin{figure}[!t]
    \centering
    \includegraphics[width=\linewidth]{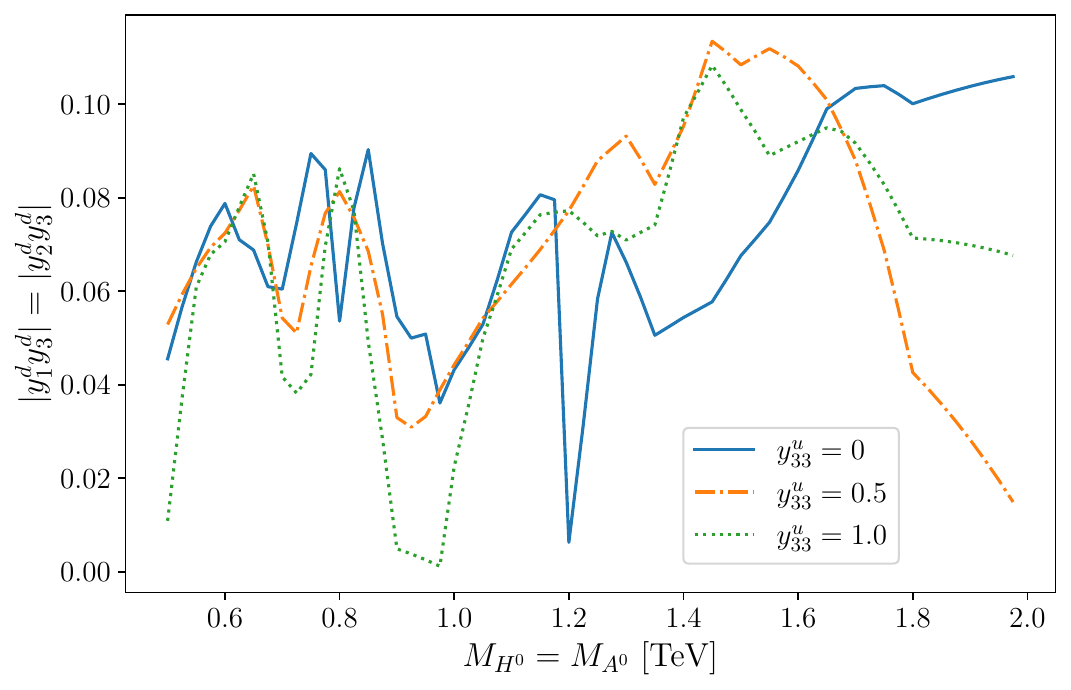}
    \caption{Effect of turning on additional $\yu{33}$ coupling on the collider bounds from the new neutral scalar component.}
    \label{fig:neutral_bounds_yu3}
\end{figure}
%%%%%%%%%%%%%%%%%%%%%%%%%%%%%%%%%%
%%%%%%%%%%%%%%%%%%%%%%%%%%%%%%%%%%
\begin{figure*}
    \centering
    \includegraphics[width=.95\linewidth]{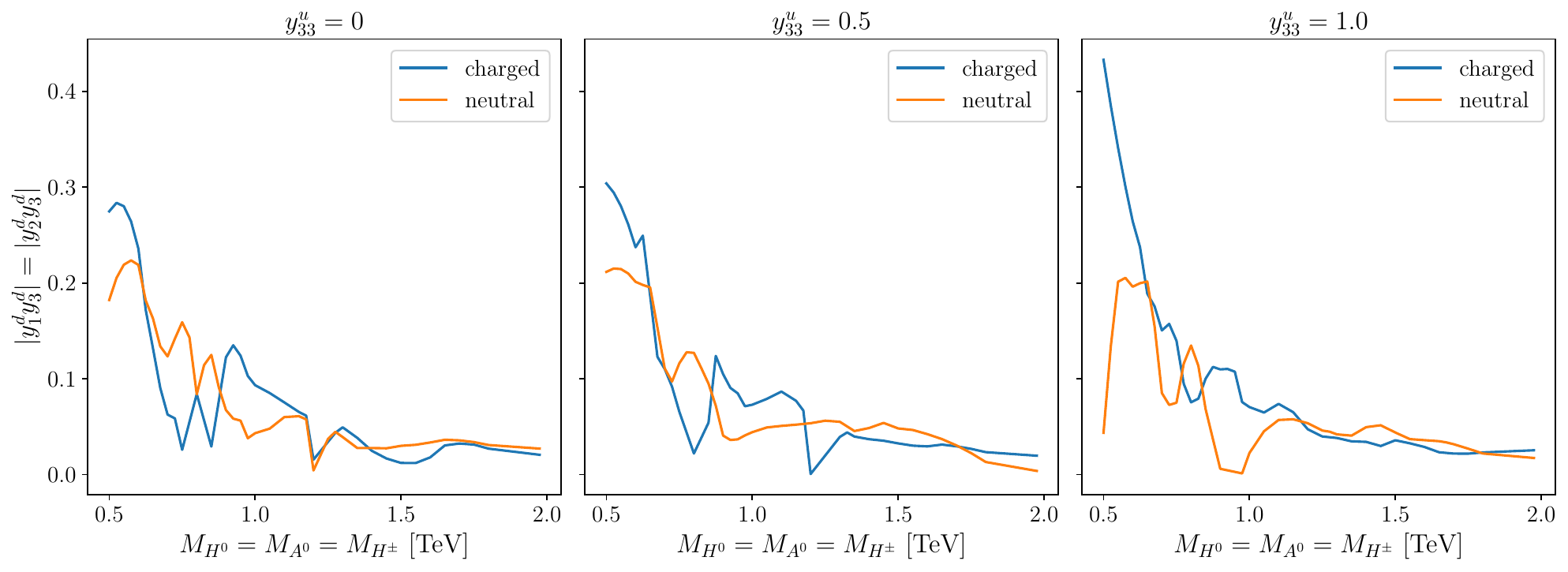}
    \caption{Comparing the collider limits from decays of the charged (blue) and neutral (orange) components of the additional doublet for $\yu{33} = 0$ (left), $\yu{33} = 0.5$ (centre), and $\yu{33} = 1$ (right).}
    \label{fig:collider_charged_vs_neutral}
\end{figure*}
%%%%%%%%%%%%%%%%%%%%%%%%%%%%%%%%%%

%%%%%%%%%%%%%%%%%%%%%%%%%%%%%%%%%%
\subsection*{Avoiding neutral component dijet bounds through \texorpdfstring{$t\bar{t}$}{tt} decays?}
%%%%%%%%%%%%%%%%%%%%%%%%%%%%%%%%%%
The non-leptonic anomaly only requires $\yd{3}$, and $\yd{1} \textrm{ or } \yd{2} \neq 0$, and turning on the additional interaction $\yu{33}$ does not directly affect these decays.
However, $\yu{33} \neq 0$ opens up a new (potentially dominantly) decay mode for the neutral component, that of $H \to t\bar{t}$, which is harder to see at LHC compared to ``bump-hunting'' dijet excesses. 
For our non-minimal model, the branching ratio of the neutral components to $t\bar{t}$ or $jj$ is given by
\begin{equation}
\begin{aligned}
\mathcal{B} (H^0 \to t \bar{t}) &\approx 1 - \mathcal{B} (H^0 \to jj)
\\
&\approx
\frac{|\yu{33}|^2 \left( 1 - \frac{4m_t^2}{M_{H^0}^2} \right)^{3/2} }{|\yu{33}|^2 \left( 1 - \frac{4m_t^2}{M_{H^0}^2} \right)^{3/2} + |\yd{3}|^2 + 2|\yd{1}|^2} \,,
\end{aligned}
\end{equation}
where we have neglected all quark masses except for the top, as well as terms proportional to $|\V{cb}|^2$ or $|\V{ub}|^2$.
We see that for $\yu{33} = 0.5$ or $\yu{33} = 1$, and the range of \yd{1,3} couplings we scan over, the branching ratio of the neutral scalar to $t\bar{t}$ will be sizeable.

While one might expect BSM to be able to hide in $t\bar{t}$ decays, we find (as shown in \cref{fig:neutral_bounds_yu3}) that in fact there is no clear effect -- in some cases the bounds get weaker, while in others they are stronger. Beyond this, comparing the best limit from charged and neutral component decay (\cref{fig:collider_charged_vs_neutral}), we also find that signals from both are roughly equally important until we get to $\yu{33} = 1$, and so even if we had been able to ``hide'' the neutral component in the QCD background of $t\bar{t}$ signals, the current charged decay limits would not have been significantly weaker. Since we already saw in \cref{fig:collider_bounds_individual} that we find stronger bounds than BGM \cite{Bordone:2021cca}, we conclude that hiding our new scalar in the top final states does not provide a viable route to reconcile the collider with flavour, when assuming QCDF provides a good description of the non-leptonic decays. Therefore, going beyond the minimal realisations considered in \cite{Bordone:2021cca} does not alleviate the data-driven tension in the combined flavour+high-energy collider data set.
In fact, the updated analysis reinforces the puzzle.

%%%%%%%%%%%%%%%%%%%%%%%%%%%%%%%%%%
\section{Underestimated QCD uncertainties vs New Physics}
\label{sec:power-corrections}
%%%%%%%%%%%%%%%%%%%%%%%%%%%%%%%%%%
As discussed in the introduction, the tensions with the SM predictions may be due not to NP effects but to underestimated QCD uncertainties. In the context of the clean observables discussed in this paper the source of these effects is however not clear -- within QCDF, the corrections to the processes $\bar{B}^0\rightarrow D^{(*)+}K^{(*)-}$ and $\bar{B}^0_s\rightarrow D^{(*)+}_s\pi^-$ beyond leading order in $\Lambda_{\rm QCD}/m_b$ are expected to be small ($\mathcal{O}(1\%)$). On the other hand, since we cannot fully address these corrections from first principles, it cannot be ruled out that these effects are larger than these first estimations. As such, the question of how large these extra purely QCD effects can really be remains.

In order to estimate reasonable bounds, we consider the situation where a $B$ meson decays into two light mesons, where power-suppressed non-factorisable contributions are prominent, as compared to the ``clean'' decays we study in this work. Using data-driven approaches, recent studies have attempted to estimate the magnitude of these effects for final states consisting of pairs of light pseudoscalar mesons \cite{Huber:2021cgk,Berthiaume:2023kmp,Davies:2024vmv,BurgosMarcos:2025xja,Fang:2026fhl}. In particular, in \cite{Fang:2026fhl}, the bounds for the power suppressed annihilation contributions $\beta_1$ and $\beta_2$ are estimated, including $SU(3)$ flavour breaking effects. We will use the results obtained for the processes $B_{(s)}\rightarrow \pi\pi$, $B_{(s)}\rightarrow K K$ and $B_{(s)}\rightarrow \pi K$ as our reference values. For these cases, the value for the power suppressed terms is placed at most at $|\beta_{1, 2}|<\mathcal{O}(20\%)$ at a $68\%$ confidence level, even in purely annihilation channels such as $B_s\rightarrow \pi\pi$.
As such, it is reasonable to expect that the power corrections in the channels we study are smaller than this.

For now, we remain agnostic to the source and analyse arbitrary universal effects, as parameterised by $\deltapc$ in \cref{eq:amplitude_with_power_corrections}. For different values of the power correction parameter $\deltapc$, the SM $\chi^2$ reduces, as does the improvement seen in our BSM scenario (see \cref{tab:bsm-power-corrections}).
%
%%%%%%%%%%%%%%%%%%%%%%%%%%%%%%%%%%
\begin{table}
    \centering
    \begin{tabular}{cccccc}
    \toprule
    \deltapc (\%) & SM $\chi^2$ & \multicolumn{3}{c}{BSM best fit} & BSM \qty{2}{\sigma} range \\
    & & $\DeltaBSM [\mathrm{TeV}^{-2}]$ &  $\Delta \chi^2$ & pull &  \\
    \midrule
    0 & 66 & -3.6 & -61 & \qty{7.8}{\sigma} & [-4.6, -2.6] \\
    -5 & 34 & -2.5 & -30 & \qty{5.5}{\sigma} & [-3.5, -1.6] \\
    -10 & 16 & -1.5 & -11 & \qty{3.3}{\sigma} & [-2.5, -0.6] \\
    -15 & 9.5 & -0.6 & -1.9 & \qty{1.4}{\sigma} & [-1.6, 0.3] \\
    \bottomrule
    \end{tabular}
    \caption{Effect of fixed universal power corrections (as parameterised by \deltapc in \cref{eq:amplitude_with_power_corrections}) on the SM $\chi^2$, the size of BSM interaction (parameterised as $\DeltaBSM$ in \cref{eq:non-leptonic_best_fit}), as well as the corresponding reduction in $\chi^2$, pull, and \qty{2}{\sigma} range of \DeltaBSM.}
    \label{tab:bsm-power-corrections}
\end{table}
%%%%%%%%%%%%%%%%%%%%%%%%%%%%%%%%%%
%
%%%%%%%%%%%%%%%%%%%%%%%%%%%%%%%%%%
\begin{figure}[!b]
    \centering
    \includegraphics[width=\linewidth]{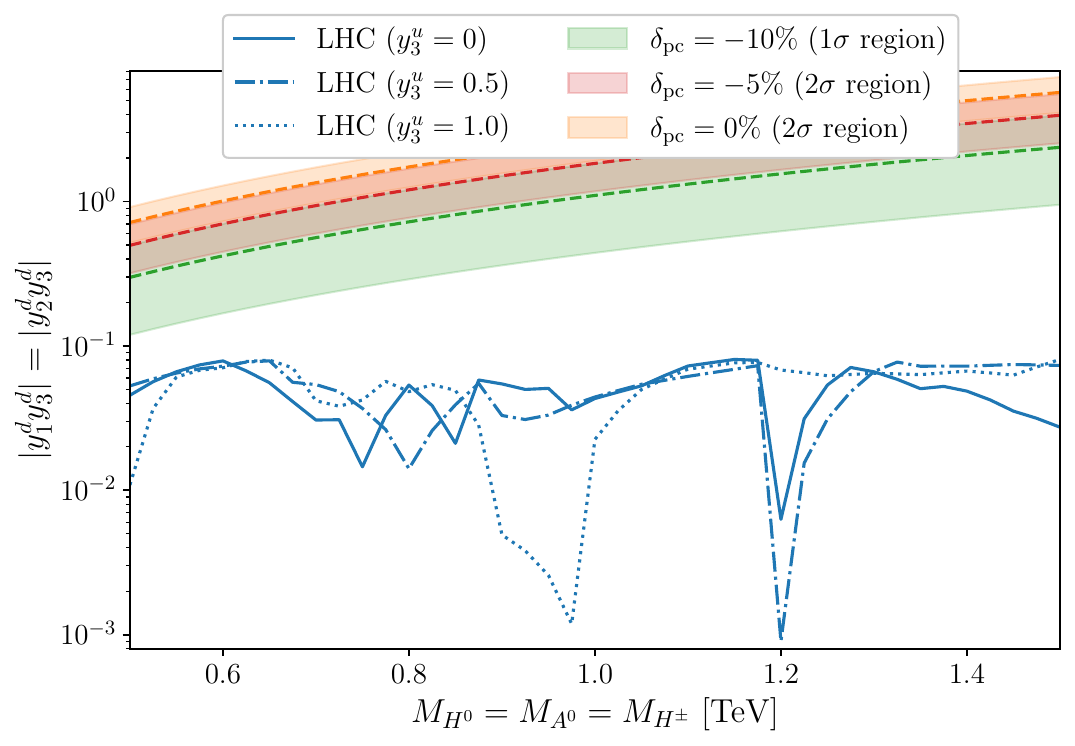}
    \caption{BSM regions for 0, 5 and 10 per cent power corrections}
    \label{fig:collider_vs_pc}
\end{figure}
%%%%%%%%%%%%%%%%%%%%%%%%%%%%%%%%%%
We find that a significant discrepancy persists, which our 2HDM model can resolve up to power corrections of order $-10\%$. Once one exceeds this level, the tension between theory and experiment can essentially be entirely addressed by a shift of around $-15\%$ from power corrections, with our BSM model only being ``preferred'' at just over \qty{1}{\sigma} in this scenario. This finding is compatible with previous results in \cite{Huber:2016xod} and \cite{Meiser:2024zea}, but as we have just discussed, such a large effect would not agree with some reasonable expectation about the hierarchy of non-factorisable effects in different hadronic $B$ decay modes.

Given that, in the case of $-10\%$ power corrections, the required size of the BSM interaction is reduced by more than half, we consider whether such large effects can reconcile collider and flavour constraints. In \cref{fig:collider_vs_pc}, we plot the collider limits found in the previous section, combining both neutral and charged scalar decay searches, for the three different values of \yu{33} considered, alongside the region favoured by the non-leptonic observables for three different values of the universal power correction. It is clear that even for a scenario with $\deltapc = -10\%$, there is still a large tension between the BSM favoured region and that allowed by the collider.

%%%%%%%%%%%%%%%%%%%%%%%%%%%%%%%%%%
\section{Many-scalar BSM scenario}
%%%%%%%%%%%%%%%%%%%%%%%%%%%%%%%%%%
In this section, we now consider an even more non-minimal BSM sector in which more than one additional Higgs doublet is present.
In such a model (ignoring fine-tuning considerations for now), the effect at low energy will combine constructively, while collider bounds from resonance searches are ``washed out''.
Specifically, assuming $n$ copies of our additional Higgs doublet with identical couplings to those in \cref{eq:bsm_lagrangian}, the contribution to the EFT coefficients given in \cref{eq:eft_wcs} is multiplied by $n$, and so the product of couplings defined by $\DeltaBSM$ in \cref{eq:non-leptonic_best_fit} can be a factor of $n$ smaller while still explaining the non-leptonic anomaly.
%%%%%%%%%%%%%%%%%%%%%%%%%%%%%%%%%%
\begin{figure}[t!]
    \centering
    \includegraphics[width=\linewidth]{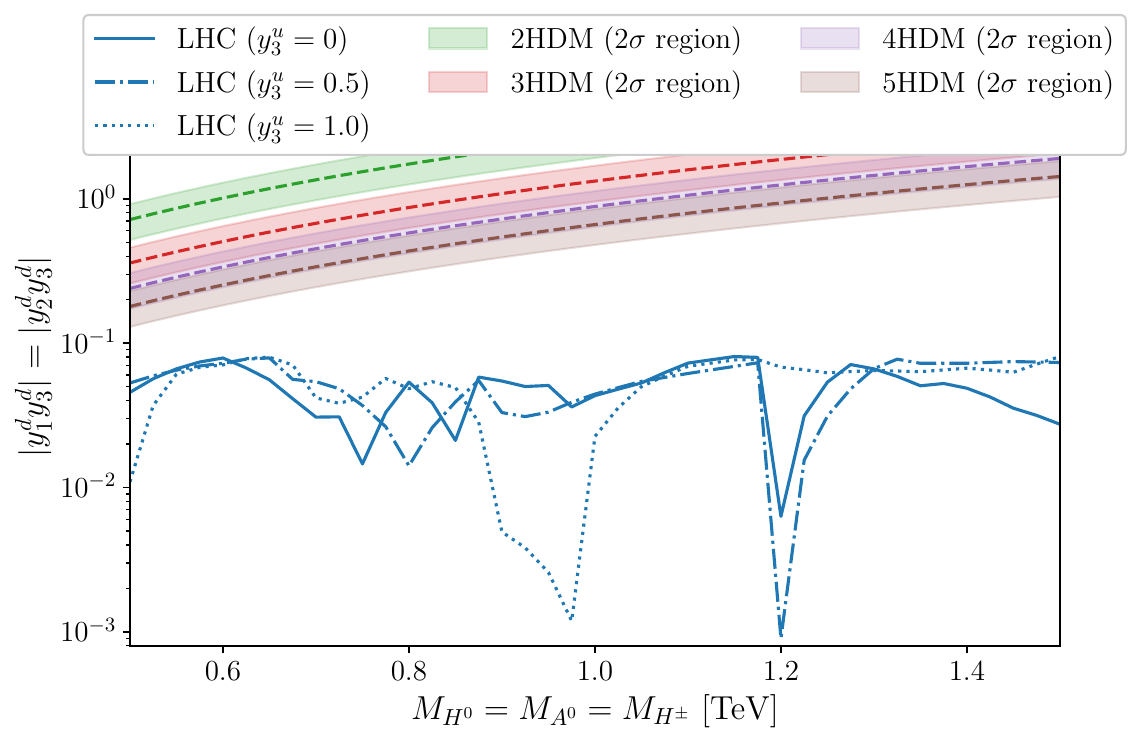}
    \caption{BSM regions for different numbers of additional doublets, each with the same BSM coupling structure.}
    \label{fig:collider_vs_nHDM}
\end{figure}
%%%%%%%%%%%%%%%%%%%%%%%%%%%%%%%%%%
%%%%%%%%%%%%%%%%%%%%%%%%%%%%%%%%%%
\begin{figure*}[!ht]
    \centering
    \includegraphics[width=\linewidth]{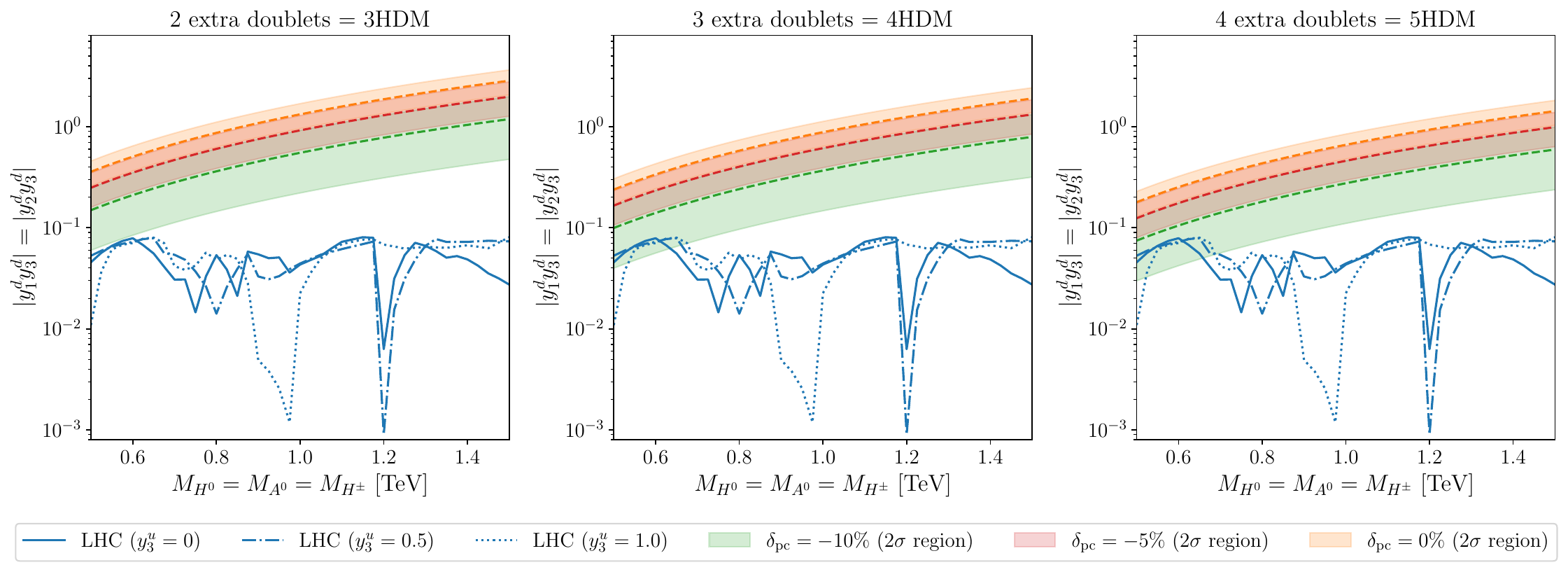}
    \caption{BSM regions for different numbers of additional doublets, each with the same BSM coupling structure, and different power correction sizes.}
    \label{fig:collider_vs_flavour_nHDM_plus_pc}
\end{figure*}
%%%%%%%%%%%%%%%%%%%%%%%%%%%%%%%%%%
As such, for a sufficiently complicated BSM sector, it is clear we are guaranteed to be able to reconcile the collider bounds with the non-leptonic anomalies. We examine extending the SM by up to four additional weak doublets, each with identical couplings to those we described in \cref{eq:bsm_lagrangian} as discussed. The corresponding BSM regions are shown in \cref{fig:collider_vs_nHDM}, where we see that even with four extra doublets (i.e. a 5HDM), a BSM scenario with no power corrections is ruled out by current collider bounds.\footnote{A quick calculation shows that a (very dubious) 14HDM would be able to explain the non-leptonic anomaly, while simultaneously being compatible with collider bounds.\label{footnote:14hdm}}

Note that in the case of degeneracy within experimental resolution, the different resonance peaks cannot necessarily be disentangled.
However, current analyses by the LHC experiments have a resolution of $\lesssim 10\%$ in dijet searches~\cite{ATLAS:2019fgd,CMS:2019gwf,ATLAS:2025okg,CMS:2025ybc}, depending on the scale of the resonance mass, and the doublet masses in our model can be straightforwardly split at this level, with a corresponding change in coupling, such that they will be experimentally resolvable without affecting any of the conclusions of our analysis.
That said, in more realistic scenarios (e.g.\ already in the 2HDM), we typically observe mass splittings between the exotic states unless additional symmetries are imposed.
In the degenerate case, the observed LHC constraints are even more restrictive as signals add, invalidating the above signal dilution.

%%%%%%%%%%%%%%%%%%%%%%%%%%%%%%%%%%
\subsection*{Many-scalar BSM + power corrections}
%%%%%%%%%%%%%%%%%%%%%%%%%%%%%%%%%%
As a final investigation, we study how all the above paths interact, i.e.\ what is possible if we allow for unexpectedly large power corrections alongside a non-trivial nHDM BSM sector.
Combining all our previous cases together, we show in \cref{fig:collider_vs_flavour_nHDM_plus_pc} what can happen if there are $-5$ or $-10\%$ power corrections for models with up to four additional scalar doublets. We see that only once we consider 4 or 5HDMs, alongside a $-10\%$ power correction, do we find a viable mass window around \qty{600}{\GeV}, and even in these cases, we are not able to fully reach the coupling size maximally preferred by the non-leptonic data.

%%%%%%%%%%%%%%%%%%%%%%%%%%%%%%%%%%
\section{Other bounds on \lowercase{n}HDMs}
%%%%%%%%%%%%%%%%%%%%%%%%%%%%%%%%%%
Beyond collider searches and non-leptonic decays, extended Higgs sectors have a rich phenomenology in the wider flavour sector, with contributions to many precision observables, for example $Z \to bb$ \cite{Bordone:2021cca,Jung:2010ik}, $B$ meson mixing~\cite{Bordone:2021cca,Jung:2010ik}, $b \to q \ell \ell$~\cite{Bordone:2021cca,Grinstein:1988me}, $b \to q \gamma$~\cite{Bordone:2021cca,Besmer:2001cj}, and the lifetime ratio of $B^+$ to $B_d$ mesons~\cite{Jager:2017gal,Jager:2019bgk,Lenz:2022pgw}.

Since our focus in this work is on the viability of the central three ideas to overcoming previously shown collider bounds, and our findings show that only somewhat contrived nHDMs with unexpectedly large power corrections beyond QCDF are viable, we provide only a minimal overview here of these issues.
We also neglect the contribution of \yu{33} in this section for simplicity, in light of the small effect that we have found $\yu{33} \neq 0$ has on the collider bounds.

Additionally, our chosen coupling structure with diagonal $y^d$ (and potentially only $\yu{33} \neq 0$) is highly specific and would, in principle, require full justification; we consider this to be beyond the scope of our study.

%%%%%%%%%%%%%%%%%%%%%%%%%%%%%%%%%%
\subsection*{$\mathbf{Z \to bb}$}
%%%%%%%%%%%%%%%%%%%%%%%%%%%%%%%%%%
Our model modifies the right-handed $Z \to b b$ interactions as
\begin{equation}
    \delta g^R_b = - \frac{(\yd{3})^2}{32\pi^2} \left( \frac{y_t^2-y_t-y_t \log y_t}{(1-y_t)^2} \right)
\end{equation}
where $y_t = m_t^2 / M_\Phi^2$ \cite{Bordone:2021cca,Jung:2010ik}.

%%%%%%%%%%%%%%%%%%%%%%%%%%%%%%%%%%
\subsection*{Low energy flavour bounds}
%%%%%%%%%%%%%%%%%%%%%%%%%%%%%%%%%%
Integrating out our additional scalar doublet at 1-loop and matching directly onto the low-energy effective theory, we find NP contributions to the following operators:
\begin{equation}
\begin{aligned}
\mathcal{L}^\textrm{BSM} &\supset \mathcal{L}^{B_q \textrm{ mixing}} + \mathcal{L}^{b \to q \ell \ell} + \mathcal{L}^{b \to q \gamma} + \mathcal{L}^{\tau(B^+)/\tau(B_d)}
\\
&= C^{VRR}_{bq} (\bar{q} \gamma^\mu P_R b)(\bar{q} \gamma_\mu P_R b) + C^{SRL}_{bq} (\bar{q} P_L b)(\bar{q} P_R b)
\\
&+ \frac{4 G_F}{\sqrt{2}} \V{tb} \V*{ts} \frac{\alpha}{4\pi} (\bar q \gamma_\mu P_R b)(\bar{\ell} \gamma^\mu (C_{9'} + \gamma^5 C_{10'}) \ell) 
\\
&+ \frac{4 G_F}{\sqrt{2}} \V{tb} \V*{ts} C_{7'\gamma}^{b \to q \gamma} \frac{e m_b}{16 \pi^2} (\bar{q} \sigma_{\mu \nu} P_L b) F^{\mu \nu}
\\
&+ \frac{4 G_F}{\sqrt{2}} \V{cb} \V*{ij} C^{ij}_{5'} (\bar{c} P_R b)(\bar{u}_i P_L d_j) \,,
\end{aligned}
\end{equation}
with the matching conditions at the EW scale:
\begin{align}
C^{VRR}_{bq} &= \frac{1}{128 \pi^2 M_\Phi^2} (\V{ts} \V*{tb})^2 (\yd{3} \yd{2})^2 y_t I_1(y_t)\,,
\\
C^{SRL}_{bq} &= \frac{1}{16 \sqrt{2} \pi^2 M_\Phi^2} G_F M_W^2 (\V{ts} \V*{tb})^2 \yd{2} \yd{3} F(y_t, x_t)\,,
\\
C^{bq\ell\ell}_{9'} &= - C^{bq\ell\ell}_{10'} = \frac{-1}{4 g^2 s_W^2} \yd{3} \yd{q} B(y_t)\,,
\\
C^{b \to q \gamma}_{7'\gamma} &= - \frac{\yd{3} \yd{q} v^2}{2 M_\Phi^2} \left( \frac{2}{3} F_1 (y_t) + F_2(y_t) \right)\,,
\\
C^{ij}_{5'} &= \frac{-\sqrt{2}}{4 G_F} \frac{\yd{3} \yd*{j}}{M_{H^+}^2}\,,
\end{align}
where $x_t = m_t^2 / M_W^2$ and the loop functions are
\begin{align}
I_1 (y) &= \frac{y+1}{(1-y)^2}+\frac{2 y \log (y)}{(1-y)^3}\,,
\\
F(y, x) &= 
    \begin{aligned}[t]
        &\frac{x(x-4)}{(x-1)(y-1)} + \frac{x^2 (4 x-7) \log x}{(x-1)^2 (x-y)}
        \\
        &- \frac{x (x+4 y (y-2) ) \log y}{(y-1)^2 (x-y)}
    \end{aligned}
\\
B(y) &= \frac{y_t \log y_t}{(y_t-1)^2}-\frac{y_t}{y_t-1}\,,
\\
F_1(y) &= \frac{x^3 - 6 x^2 + 3x + 2 + 6x \log x}{12 (x-1)^4} \,,
\\
F_2(y) &= \frac{2 x^3 + 3x^2 - 6x + 1 - 6x^2 \log x}{12 (x-1)^4} \,.
\end{align}

%%%%%%%%%%%%%%%%%%%%%%%%%%%%%%%%%%
\subsection*{Benchmark assessment}
%%%%%%%%%%%%%%%%%%%%%%%%%%%%%%%%%%
We take a benchmark point of $M_\Phi = \qty{600}{\GeV}$, since at this mass the collider bounds come closest to the NP interaction needed to explain the non-leptonic anomalies. Using the global likelihood provided by \smelli \version{2.4.2} \cite{Aebischer:2018iyb,Stangl:2020lbh} (as well as SLC and LEP results for the $Z \to bb$ couplings \cite{ALEPH:2005ab}, and HFLAV average of $\tau(B^+)/\tau(B_d)$ \cite{HeavyFlavorAveragingGroupHFLAV:2024ctg,HFLAV:2024osc}), we find that the parameter space not ruled out by LHC searches is also not in tension with the $Z$ pole or low energy bounds, with the partial exception of $\Delta M_s$,\footnote{The non-perturbative input for $B$ meson mixing corresponds to the combination in \cite{Greljo:2022jac} of \cite{FermilabLattice:2016ipl,Dowdall:2019bea,Kirk:2017juj,King:2019lal}.} where a tension of up to \qty{3}{\sigma} with the experimental measurements \cite{HeavyFlavorAveragingGroupHFLAV:2024ctg} can be present, depending on the relative sign of \yd{3} and \yd{2}, which could wash out the improvement in a global fit from explaining the non-leptonic anomalies. Alternatively, since $\Delta M_d$ is much less constraining, our model could avoid tension with other flavour measurements by only coupling to first and third generation quarks, at the expense of reducing its explanatory power in the non-leptonic sector.

%%%%%%%%%%%%%%%%%%%%%%%%%%%%%%%%%%
\section{Conclusions}
%%%%%%%%%%%%%%%%%%%%%%%%%%%%%%%%%%
For many years now, there has been an extremely puzzling discrepancy seen in $\bar{B}^0\rightarrow D^{(*)+}K^-$ and $\bar{B}^0_s\rightarrow D^{(*)+}_s\pi^-$ decays, when comparing the experimentally measured branching ratios to the clean predictions given in the QCDF framework.
Previous studies have produced many viable BSM explanations in terms of BSM EFT coefficients for four-quark effective operators, but attempts to construct UV completions have been hindered by the connection to dijet production at the LHC.\footnote{We again note that a 2HDM with the second doublet having a mass very close to the top quark was provisionally identified in Ref.~\cite{Bordone:2021cca} but was not fully explored there and so, in principle, remains potentially viable.}
In this article, we have investigated three potential options to increase the viability of a 2HDM as an explanation of the non-leptonic decay anomalies, namely
\begin{enumerate}
    \item that NP is hiding in discovery modes that are difficult to constrain, specifically through top-philic components,
    \item that QCD is not as well under control as we believe and creates tension in the $B$ sector through unexpectedly large power corrections,
    \item that the NP sector contains multiple new particles, which dilutes collider bounds,
\end{enumerate}
all of which, a priori, are reasonable and plausible.

Firstly, we demonstrated that, despite the substantial interference and background effects in searches for $t\bar{t}$ final states, our reanalysis of dijet searches remains sensitive to the neutral component, even with a large branching ratio to that decay mode. Beyond this, the charged component decays are approximately equally as strong, and so even if the signs of the neutral doublet component had been obscured, evidence from the charged scalar decays would also provide similar limits on BSM couplings.

Next, we discussed the potential size of power corrections in the leading-order QCDF expansion.
Recent work comparing theory and data for non-leptonic decays with large ``pollution'' from non-factorisable contributions such as $B_{(s)}\rightarrow \pi\pi$, $B_{(s)}\rightarrow K K$ and $B_{(s)}\rightarrow \pi K$ suggests a bound for the power corrections of $20\%$.
We find that our decays, which are in principle much less sensitive to these effects, would require at least $15\%$ power corrections to explain the data, and so this, in and of itself, would be a somewhat unexpected result if a first-principles calculation were possible.
Proceeding agnostically, we showed that, for somewhat smaller power corrections that apply universally to all decay modes, the discrepancy remains.
In the $-10\%$ power correction scenario, our BSM model continues to be favoured by more than \qty{3}{\sigma}, but with the size of NP coupling reduced by more than a factor of two. However, even with this smaller coupling, we find that the limits from dijet data remain too strong.

Our third avenue was to go beyond the 2HDM by allowing additional copies of the additional scalar doublet with identical coupling structures. The additional scalars all combine constructively in the low-energy non-leptonic sector, thus requiring a smaller NP coupling per field, offering the possibility to drop below the collider searches for coloured resonances. Considering up to four additional scalar doublets, all with the same BSM coupling structure, we find that there remains a large gap between the required interaction strength to reconcile the non-leptonic decay data with theoretical predictions and what would remain unseen in current collider data.

We finally studied the interplay among these ideas, considering a non-trivial BSM scenario with multiple additional Higgs doublets, in which the neutral component is potentially top-philic and large-power corrections affect the SM prediction. Only here do we find a (naively) viable model with two additional scalar doublets, degenerate in mass and interaction structure, at around \qty{600}{\GeV}, capable of (partially) resolving the non-leptonic anomalies.

Of course, there are further flavour constraints on extended scalar sectors, which we have assessed. Studying a benchmark point that is not ruled out by our collider bounds, we find that $B_s$ mixing provides the next strongest constraint, potentially negating the overall fit improvement from the reduction in tension in the non-leptonics (assuming we wish to explain the tensions in both $b \to c u d$ and $b \to c u s$ decays).

In summary, we find that the anomalies seen in $\bar{B}^0\rightarrow D^{(*)+}K^{(*)-}$ and $\bar{B}^0_s\rightarrow D^{(*)+}_s\pi^-$ decays are particularly difficult to explain through NP models, given the strong expected signals from colliders, with various typical model building routes failing to allow construction of a viable model except in particularly tuned scenarios, which include large power corrections beyond what would be naively expected from QCDF.
As such, we expect that these discrepancies will remain a puzzle for the foreseeable future.

%%%%%%%%%%%%%%%%%%%%%%%%%%%%%%%%%%
\section*{Acknowledgements}
%%%%%%%%%%%%%%%%%%%%%%%%%%%%%%%%%%
We thank Stefan Meiser for discussions regarding the power correction analysis in Ref.\ \cite{Meiser:2024zea}. M.K.\ thanks the participants of the IPPP ``Nonleptonic Decays of Heavy Mesons'' workshop for useful discussions, and to whom Footnote~\ref{footnote:14hdm} is dedicated.

C.E.\ and J.Y.A.\ are supported by the Institute for Particle Physics Phenomenology Associateship Scheme. G.T.-X. is supported by the European Union’s Horizon 2020 research and innovation program under the Marie Sk{\l}odowska-Curie grant agreement No 945422. This research was supported by the Deutsche Forschungsgemeinschaft (DFG, German Research Foundation) under grant 396021762 - TRR 257. JYA is supported by the Royal Society under grant no. IES/R2/252139.

%%%%%%%%%%%%%%%%%%%%%%%%%%%%%%%%%%
\appendix
%%%%%%%%%%%%%%%%%%%%%%%%%%%%%%%%%%
%%%%%%%%%%%%%%%%%%%%%%%%%%%%%%%%%%
\section{Collider uncertainties}
\label{sec:app}
%%%%%%%%%%%%%%%%%%%%%%%%%%%%%%%%%%
The inclusion of the full statistical model in the reinterpretation framework has been repeatedly demonstrated to be an essential component of the reinterpretation pipeline.  Simplified approaches that rely solely on observed upper limits on the production cross-section neglect the shape of the underlying likelihood, including systematic uncertainties and their correlations, which can lead to materially different conclusions on the excluded parameter space. In order to quantify this effect on our exclusion estimates, we adopt an asymptotic approximation in which the interpolated cross-section limits are used to assign a likelihood to each signal hypothesis at a given mass point,
%%%%%%%%%%%%%%%%%%%%%%%%%%%%%%%%%%
\begin{figure}
    \centering
    \includegraphics[width=0.9\linewidth]{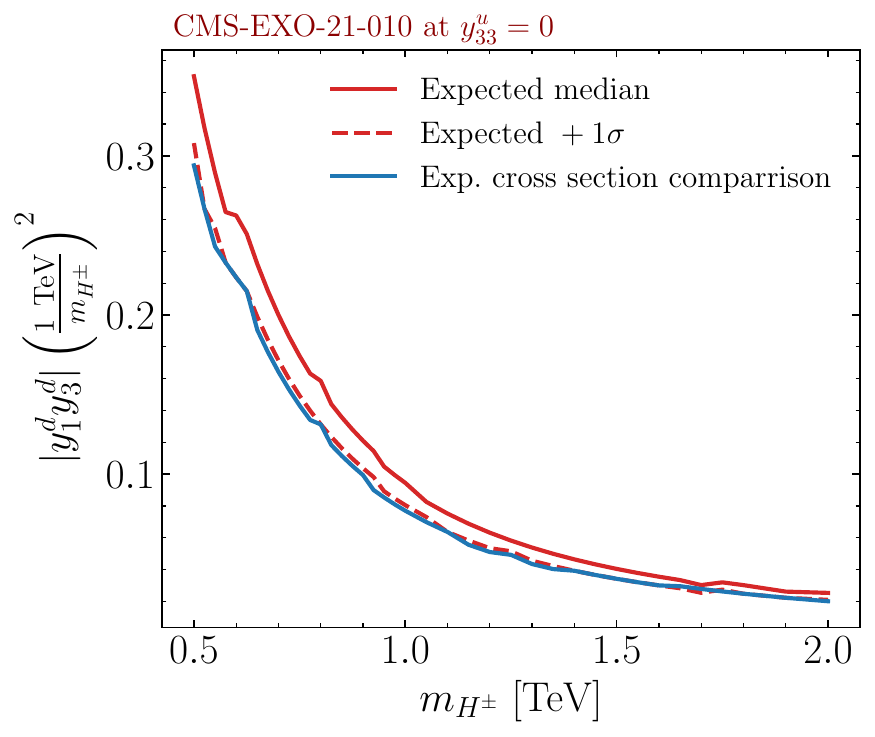}
    \caption{Comparison of exclusion limits on the Yukawa couplings $|y^d_1 y^d_3|$ as a function of the charged Higgs mass $m_{H^\pm}$, evaluated for the CMS-EXO-21-010 dijet analysis at $y^u_{33} = 0$. The blue curve shows the naive cross-section comparison, whilst the red curve shows the expected median exclusion at 95\% $\text{CL}_s$ computed via \textsc{Spey}, with the red dashed line indicating the $+1\sigma$ uncertainty on the expected limit. \label{fig:exclusion_unc}}
\end{figure}
%%%%%%%%%%%%%%%%%%%%%%%%%%%%%%%%%%
%
\begin{equation}
    \mathcal{L}(\mu) = \frac{1}{\sigma\sqrt{2\pi}}
    \exp\!\left[-\frac{1}{2}\left(
    \frac{\mu\, \sigma_{\text{sig}} + \sigma_{\text{exp}} - \sigma_{\text{obs}}}{\delta\sigma}
    \right)^{\!2}\right],
\end{equation}
where $\sigma_{\text{sig}}$ denotes the BSM signal cross-section times branching fraction times acceptance ($\sigma \times B \times A$), $\sigma_{\text{exp}}$ is the expected (SM-only) upper limit, $\sigma_{\text{obs}}$ is the observed upper limit, and $\delta\sigma$ is the $1\sigma$ uncertainty on the expected limit. This construction treats the ratio of the signal-plus-background prediction to the observed result as a normally distributed test statistic, with the experimental uncertainty $\delta\sigma$ encoding the analysis' sensitivity at each mass point. The signal strength modifier $\mu$ is then profiled to determine the cross-section value excluded at a given confidence level.

Using \textsc{Spey} (version~0.2.6)~\cite{Araz:2023bwx} and its built-in exclusion limit calculator, we estimated the maximum signal cross-section consistent with a $\text{CL}_s$ exclusion at 95\% confidence level, together with the $1\sigma$ uncertainty band on this limit arising from the experimental sensitivity.
Figure~\ref{fig:exclusion_unc} shows the comparison between the cross-section presented in the main analysis (blue) and the expected exclusion limits with their $1\sigma$ uncertainty envelope (red) for the CMS-EXO-21-010 dijet analysis, evaluated at $y^u_{33} = 0$ across the charged Higgs mass range $500 \lesssim m_{H^\pm} \lesssim 2000~\text{GeV}$. We consistently observe discrepancies of up to $1\sigma$ between the two approaches, with the naive cross-section comparison systematically overestimating the excluded coupling strength compared to the proper statistical treatment. The results confirm that incorporating the full available statistical information, including the uncertainty on the expected limit, is necessary for reliable exclusion contours in BSM reinterpretation studies. It is important to note that this approach only naively represents the statistical model; for a systematic analysis of the effects of the uncertainties, full statistical models are necessary, along with the analysis logic.

%%%%%%%%%%%%%%%%%%%%%%%%%%%%%%%%%%
\bibliographystyle{apsrev4-1}
\bibliography{references} 
%\bibliography{paper.bbl}
%%%%%%%%%%%%%%%%%%%%%%%%%%%%%%%%%%
\end{document}